# Scaling a CUDA-Q GQE + QSCI pipeline to 40 qubits for EUV photoresist chemistry

Karim Elgammal [1] and Marc Maußner [2]

[1] RISE Research Institutes of Sweden [2] infoteam Software AG

Correspondence: karim.elgammal@ri.se

**Abstract**

We scale a CUDA-Q-native pipeline coupling a generative quantum eigensolver (GQE) to quantum-selected configuration interaction (QSCI) across active spaces of 14 to 44 qubits, applied to the extreme-ultraviolet (EUV) photoresist chemistry of monoalkyltin oxo-hydroxides. A GPT-2 policy emits UCCSD operator sequences; sampled bitstrings become determinants, diagonalised classically, and a cross-circuit generalised-eigenvalue refinement makes every reported GQE+QSCI energy a variational upper bound. Every rung from 14 to 40 qubits carries an exact CASCI or FCI reference, up to 166 million determinants for SnO at 32 qubits. The pipeline is chemically accurate, below 1.6 mHa, through 30 qubits on methyltin trihydroxide and through 32 on SnO, on the best seed at the top rungs. Circuit depth rather than training length is the scaling lever; the refined subspace grows near-linearly with the operator count while staying a vanishing fraction of the determinant space, 0.017% at the 32-qubit SnO rung. It also runs on the 54-qubit IQM Emerald processor, at the shallow depths its routed two-qubit gates allow, reaching +0.330 mHa for SnO at 14 qubits from a CCSD-amplitude-ordered pool prefix and +3.92 mHa for the industrial n-butyltin ligand at 22 qubits from depth-truncated trained circuits under per-circuit readout self-calibration, 81% of the active-space correlation; classical configuration recovery on those counts tightens the 22-qubit result to +0.18 to 0.21 mHa. For the methyl resist, ionisation collapses the classical UCCSD(T) Sn-C bond dissociation energy from 72.6 to 21.2 kcal/mol, the switch that flips solubility on exposure. Against that, the 40-qubit result is support-limited at 22.8 mHa, the full trained ansatz on hardware awaits better fidelities, and classical subspace expansion reaches the 32 to 40-qubit spaces with no quantum sampler, so that boundary is mapped, not beaten.

## 1. Introduction

Extreme-ultraviolet (EUV) lithography at 13.5 nm is the critical patterning step in advanced semiconductor manufacturing [1]; photoresist chemistry is its central materials bottleneck. The commercialised monoalkyltin oxo-hydroxide (RSnOOH) resist platform [2,3] works through one reaction: EUV exposure ionises the resist and the weakened tin-carbon bond then breaks, flipping solubility [4–6]. That switch is multireference, bond-breaking chemistry, exactly where single-determinant density functional theory (DFT) fails [7] and correlated wavefunction methods scale exponentially [8,9]. This study simulates that chemistry directly with the model monomer methyltin trihydroxide $CH_3Sn(OH)_3$ [4,10] and the real industrial ligand, n-butyltin trihydroxide [11,12], in active spaces from 22 to 44 qubits. The SnO fragment [13] provides a ladder of smaller reference systems whose energies can still be checked exactly.

We build on Kemmoku et al. [14]. To it we add an open-shell QSCI port (doublets), a measured depth-as-coverage scaling law with its GEVP-subspace mechanism [15], per-circuit readout self-calibration on real hardware (each circuit measures its own map with no prior device characterisation), S-CORE classical-boundary mapping [16,17], and a persistent multi-GPU container sampling bridge. The generative model itself contributes circuits that no classical selected-CI method [18] emits, shallow enough to run on today's hardware once depth-truncated, produced by one unchanged training protocol across all 18 homolysis and cation-curve geometries.

The Methods section sets out the pipeline and the platforms it runs on. The Results section reports the qubit ladder, the circuit-depth lever, the multi-node scaling, the chemistry of the exposure switch and the hardware runs, each benchmarked against an exact reference wherever one exists. The Discussion analyses the accuracy bottleneck, the role of device noise, scaling, the classical baselines and how to reproduce the results.

## 2. Methods

We combine the generative quantum eigensolver (GQE) [19,20], which trains a classical model to emit good circuits, with quantum-selected configuration interaction (QSCI, also called SQD) [16,21], which samples bitstrings from those circuits, selects determinants and diagonalises the Hamiltonian classically in the sampled subspace (Figure 1). Our stack adopts the work of Kemmoku et al. [14] on CUDA-Q [22]. A 6-layer GPT-2 policy [23] (43 to 46M parameters, with the operator-pool vocabulary) autoregressively emits sequences of $L$ operators from a UCCSD pool of ansatz screened by CCSD amplitudes [24]. Each sequence is a circuit of `exp_pauli` evolutions on the Hartree-Fock reference, sampled with `cudaq.sample` (up to $10^5$ shots/circuit in simulation, $10^4$ at the 44-qubit rung, 5000 on hardware). Bitstrings are mapped to determinants, post-selected by particle number, completed under total-spin symmetry, and diagonalised (PyCI [25]). A generalised-eigenvalue-problem (GEVP) refinement [15] merges the determinant spaces of all circuits, so every GQE+QSCI energy reported here is a variational upper bound. The ENPT2 tail column is a perturbative estimate, not a bound. An exponential logit-matching loss (learning rate $5 \times 10^{-7}$, replay buffer 200) updates the training policy. An A/B study against the repository's GRPO loss [26] shows the result is loss-robust (both reach chemical accuracy on SnO at 14 and 18 qubits, GEVP error 0.000 mHa in all eight runs). An open-shell port (ROHF references [27], unrestricted amplitude screening [24], Jordan-Wigner generators [28]) treats doublets, validated on the OH radical [29] to $10^{-11}$ mHa of exact CASCI.

Part of the training and simulation ran on EuroHPC Leonardo (NVIDIA A100 64 GB, 4 GPUs/node, CUDA-Q 0.14, reproducible `pixi` environment); real-hardware QSCI ran on the 54-qubit IQM Emerald QPU via CUDA-Q's native `iqm` target, every circuit is validated in emulation before it is submitted to the device. The same workflow runs on qBraid using on-demand NVIDIA A100 80 GB instances, and the qBraid runtime access to the IQM Emerald QPU as well as the QIR simulator on which the full hardware protocol was validated.

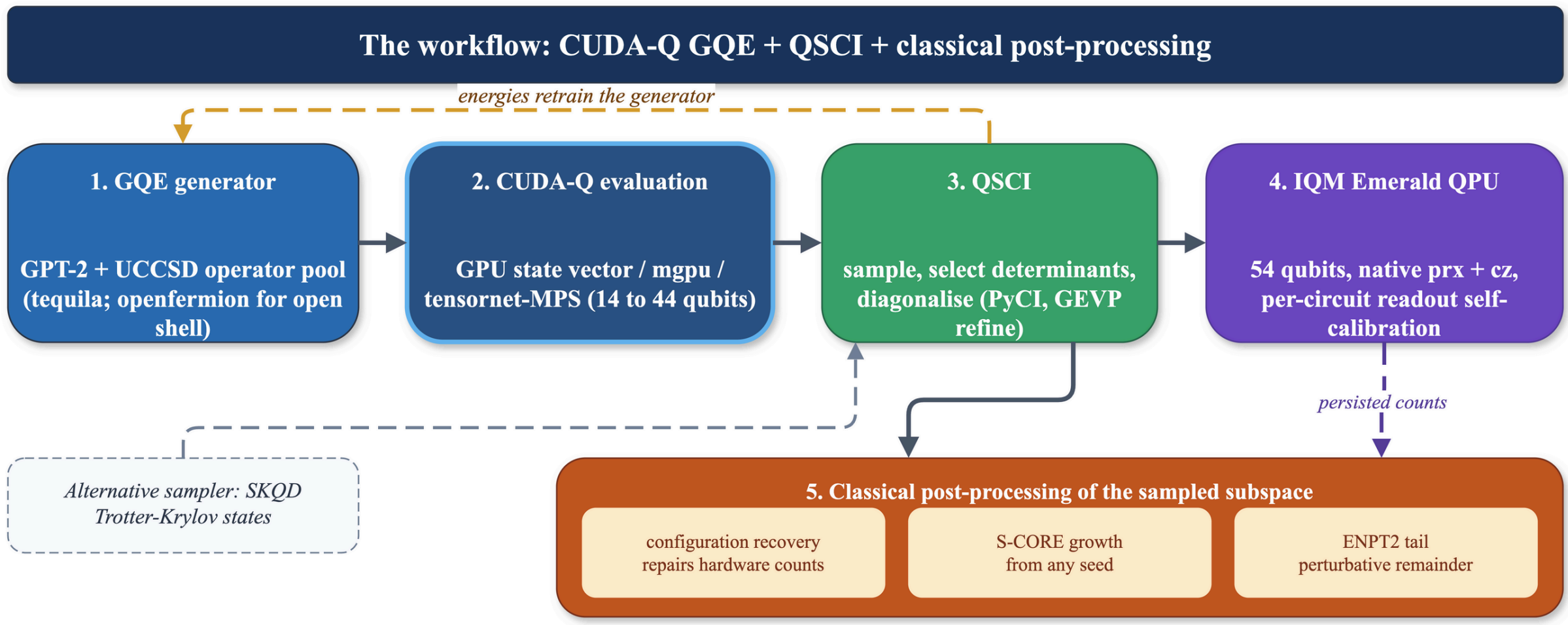

Figure 1: The hybrid system. GPT-2 policy (GPU) → CUDA-Q circuit sampling (single-GPU fp32 statevector to 32q; pooled 4-GPU `mgpu` at 34q via a persistent containerised worker; `tensornet-mps` at 40q; real IQM Emerald via the native `iqm` target) → classical QSCI diagonalisation and GEVP refinement → loss update, with classical post-processing of the sampled subspace (details in Table 1).

The target systems are the industrial monoalkyltin EUV photoresists $CH_3Sn(OH)_3$ and n-$BuSn(OH)_3$ (n-butyl), together with the SnO validation ladder, in active spaces of 14 to 44 qubits. The resist spaces are selected automatically by atomic valence active-space projection (AVAS) [30], the SnO ladder by a valence energy window. The transformer models the circuit-sequence distribution over the UCCSD pool, and its training data is the online replay buffer of (sequence, QSCI-energy) pairs. About 28,000 GPU-hours were consumed, and costs are reported per rung, one rung being one active-space size. Sn-C bond homolysis is the EUV exposure switch in candidate resists for sub-2 nm lithography, and the ionisation-collapsed bond strengths and ligand-dependent fragmentation channels quantified here are the property gaps classical methods leave at these active-space sizes.

## 3. Results

Accuracy was measured as a function of active-space size. The main results are reported in Table 1 with one rung per row. From the SnO fragment at 14 qubits to the full resist molecule at 40, the pipeline trains at every rung, each with an exact CASCI/FCI reference [31], up to a 166M-determinant FCI at SnO 32q. Every reported error is against an exact solution.

| System | Active space | Qubits | Dets | GQE+QSCI | CA | Subspace | + ENPT2 | SKQD | S-CORE (classical) |
|---|---|---|---|---|---|---|---|---|---|
| SnO | (2e,7o) | 14 | 49 | 0.000 | yes | - | - | 0.000 | - |
| SnO | (2e,9o) | 18 | 81 | 0.000 | yes | - | - | 0.000 | - |
| SnO | (10e,8o) | 16 | 3136 | 0.12 / 0.10 | yes | 34 | +0.002 / +0.007 | 0.27 | - |
| SnO | (12e,10o) | 20 | 44.1k | 0.73 / 0.73 | yes | 5.1 | - | 0.47 | - |
| SnO | (14e,12o) | 24 | 627k | 0.31 ($L$=80) | yes | 1.9 | - | 0.90 | - |
| SnO | (14e,14o) | 28 | 11.8M | 0.98 ($L$=80) | yes | 0.24 | - | 1.96 | - |
| SnO | (16e,16o) | 32 | 166M | 1.45 ($L$=80) | yes | 0.017 | - | 4.37 | - |
| resist | (14e,11o) | 22 | 108.9k | 1.42 / 1.41 | yes | 0.88 | - | 0.30 | - |
| resist | (14e,13o) | 26 | 2.9M | 1.38 / 1.03 ($L$=600) | yes | 1.9 | - | 0.88 | - |
| resist | (24e,15o) | 30 | 207k | 1.21 ($L$=600) | yes | 13 | - | 1.63 | - |
| resist | (24e,16o) | 32 | 3.3M | 2.60 / 3.64 ($L$=800) | no | 2.6 | −0.06 / −0.09 | 5.62 | 0.58 / 0.58 seeded; 1.48 from HF |
| resist | (26e,17o) | 34 | 5.66M | 4.55 / 4.93 ($L$=600) | no | 0.82 | −0.06 / −0.06 | - | 0.83 / 0.84 seeded |
| resist | (32e,20o) | 40 | 23.5M | 22.8 ($L$=10) | no | 0.013 | +0.64 | - | 1.43 seeded; 1.25 from HF |
| resist | (32e,22o) | 44 | 5.6G | - | - | 0.00004 | - | - | - |
| resist cation | (13e,11o) | 22 | 152.5k | 1.08 ($L$=80) | yes | 11 | - | - | - |
| n-Bu resist | (14e,11o) | 22 | 108.9k | 0.72 / 0.82 | yes | 0.92 | - | 0.35 | - |
| n-Bu cation | (13e,11o) | 22 | 152.5k | 1.21 / 1.22 | yes | 0.93 | - | - | - |

Table 1: The trained qubit ladder, with three alternative routes on the same Hamiltonians. Errors are in mHa against each rung's exact CASCI/FCI [31]; a dash marks a combination not run. The "resist" is $CH_3Sn(OH)_3$ with AVAS-selected active spaces [30]; n-Bu rows are the industrial ligand n-$BuSn(OH)_3$ in the same descriptor spaces (all three depicted in Figure 2); cation rows are open-shell doublets. GQE+QSCI is the GEVP-refined error, CA marks chemical accuracy, below 1.6 mHa, and Subspace is the refined subspace at that result as a percentage of the determinant space. It is dashed at 14 and 18 qubits, where refinement is exact but the subspace was not recorded. Two values are seeds 42 / 137; one is best-seed (the SnO and resist-cation $L$ = 80 rungs; both SnO seeds chemically accurate at 24q and 28q) or single-seed (resist 30q $L$ = 600, seed 137; 40q, seed 42). Depth $L$ = 10 unless noted; 40q and 44q use CUDA-Q's `tensornet-mps` backend, all other rungs full statevectors. 200 training iterations (resist 26q and 32q to 600, 30q to 400, 34q to 365, 40q to 113). The Epstein-Nesbet tail (screening $\varepsilon = 10^{-5}$) estimates the remainder rather than bounding it, so a negative entry lies below the exact energy. Resist active spaces are tinted by region, matching Figure 2: red where AVAS selects only the Sn-C valence, blue where the O 2p shells bring the Sn-O crosslinking network in; the SnO rungs use a valence energy window, not AVAS, so they are untinted, and rung spacing follows that chemistry rather than uniform qubit steps, which is why there is no 28-qubit rung. SKQD and S-CORE are defined in the Discussion; S-CORE runs use different determinant budgets and are not matched to one another. The 44-qubit row carries no error: no exact reference could be computed.

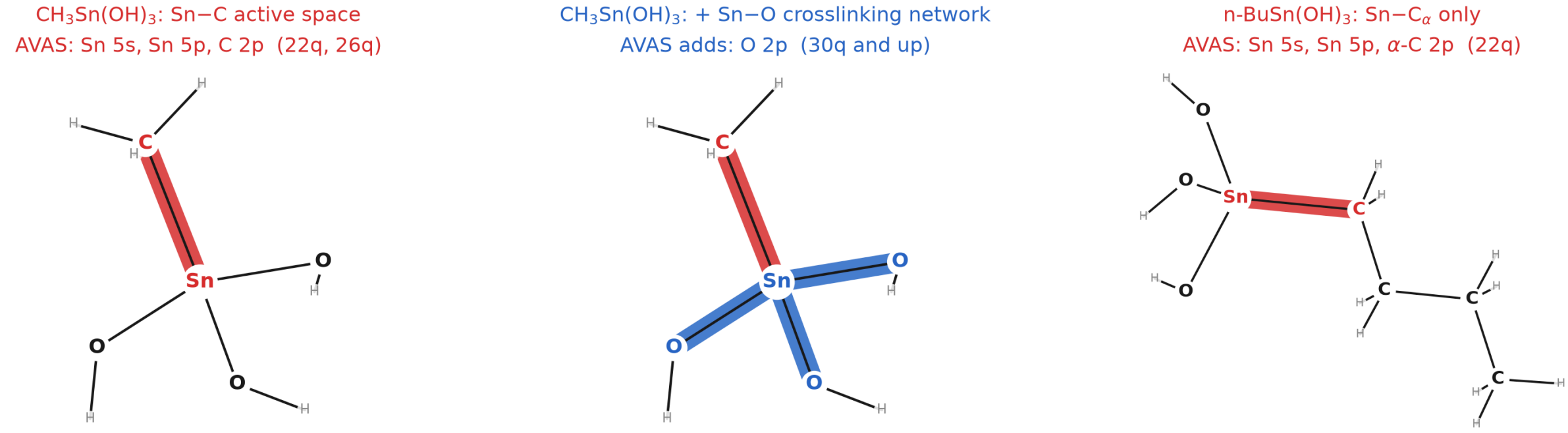

Figure 2: The AVAS active-space regions, tinted as in Table 1. Left and centre, the resist $CH_3Sn(OH)_3$: the Sn-C space, the EUV solubility switch (AVAS Sn 5s, Sn 5p, C 2p; 22q and 26q), then the added Sn-O crosslinking network (+O 2p; 30q and up). Right, the industrial ligand n-$BuSn(OH)_3$: its AVAS label is atom-indexed on the alpha carbon, so the rest of the chain stays outside the active space and the ligand occupies the same (14e,11o) space at 22q as the methyl resist, the one rung at which it is run.

Circuit depth is the scaling driver, one of two tested at fixed budget (200 iterations, determinant cap $10^5$). Longer training saturates and by extending the shallow resist runs from 200 to 600 iterations, it improves rungs by 25 to 37%. Longer generated circuits do close the gap. Raising $L$ from 10 to 80 carries the SnO ladder to chemical accuracy at 24q (3.36 → 0.31 mHa), 28q (8.64 → 0.98) and 32q (10.06 → 1.45). $L = 600$, the several-hundred-operator regime of the base framework [14], carries the resist across at 26q on both seeds (1.38/1.03 mHa) and at 30q (1.21 mHa, crossing 1.6 at iteration 204 of 400). We measured why depth works as the GEVP subspace grows near-linearly with $L$ yet stays a vanishing fraction (0.017% of the 166M-determinant 32q space), so depth buys coverage without classical enumeration (Figure 3). A richer operator vocabulary instead hurts.

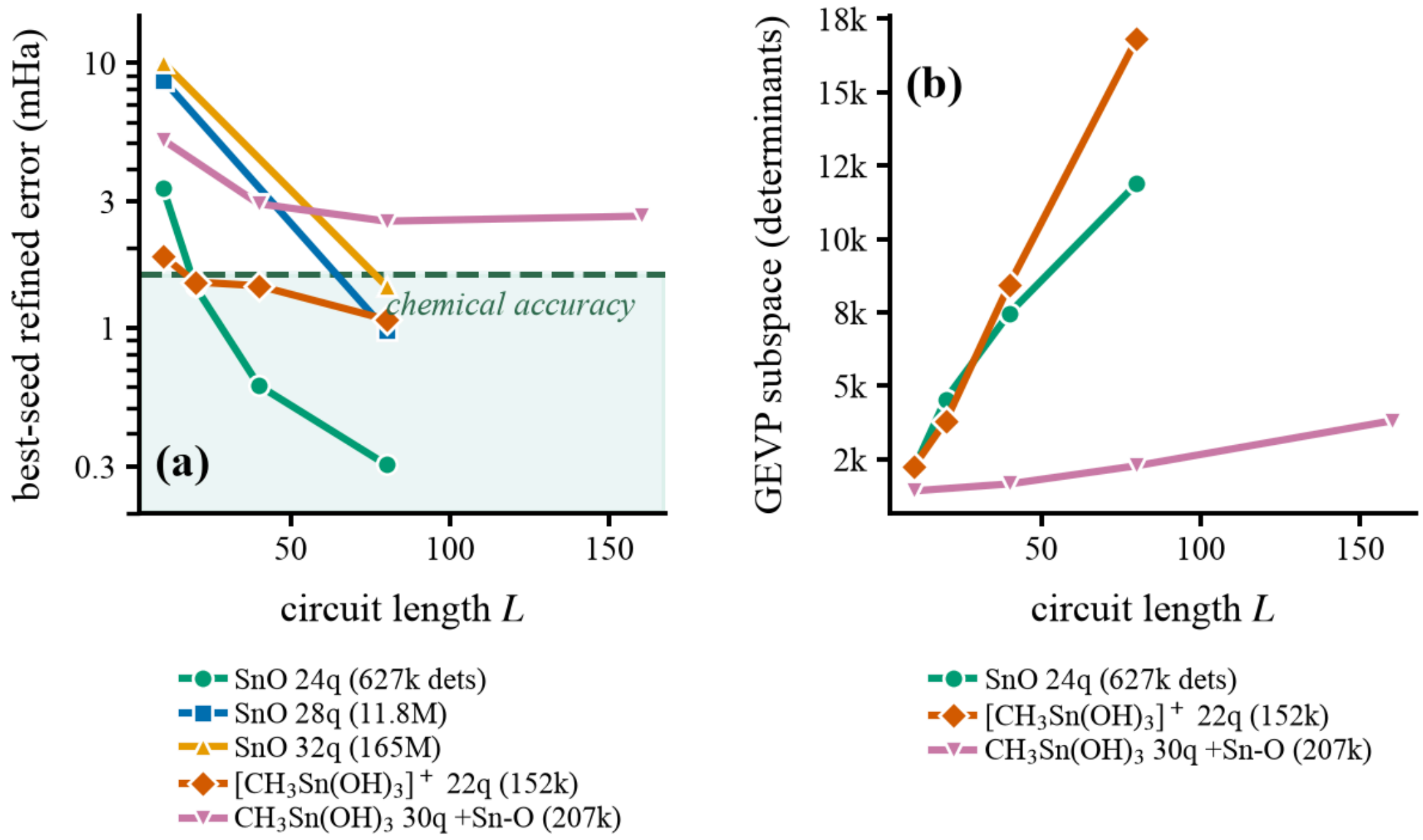

Figure 3: Circuit depth closes the scaling gap (best seed, fixed 200-iteration budget and determinant cap; GEVP-refined error against each system's exact CASCI). (a) Error falls with the generated-operator count $L$: the SnO 24, 28 and 32q rungs enter the chemical-accuracy band (shaded) from 3.36, 8.64 and 10.06 mHa at $L = 10$. (b) Mechanism: the cross-circuit GEVP subspace grows near-linearly with $L$; the resist's concentrated 24-electron wavefunction grows an order of magnitude more slowly, the origin of its saturation (broken at $L = 600$, see text).

| Qubits | Nodes | GPUs | s / 5000-shot sample |
|---|---|---|---|
| 34 | 1 | 4 | 2.85 |
| 35 | 2 | 8 | 2.83 |
| 36 | 4 | 16 | 20.4 |
| 37 | 8 | 32 | 73.0 |
| 38 | 16 | 64 | 70.3 |
| 39 | 32 | 128 | 90.2 |
| 40 | 64 | 256 | 107.2 |

Table 2: Measured multi-node exact-statevector weak scaling on Leonardo A100 64GB (containerised CUDA-Q `mgpu`, GQE `exp_pauli` kernel, $L = 32$, 5000 shots, fp32; per-GPU slice constant at 34.4 GB).

At 32 qubits the state needs 64 GiB in fp64, which fills one Leonardo HPC A100, but only 32 GiB in fp32, which fits, so 32q training uses the full statevector with nothing truncated. Lower precision changes only the sampled bitstrings; the diagonalisation and all references stay fp64. At 34 qubits one GPU is too small, so we pool four with `mgpu` inside a container, and a persistent worker holds that container open so it no longer restarts on every rollout. Training at $L = 600$ then reaches 4.55/4.93 mHa by iteration 365, half the earlier shallow single-seed result. At 40 qubits no statevector fits, so an MPS sampler with a bounded bond dimension replaces it. One circuit takes 2.8 s on a single A100, and raising the depth from $L = 20$ to 40 costs almost nothing at $\chi = 64$. A 100,000-shot rollout takes about 73 min, most of it drawing shots rather than contracting tensors. Against the exact 23.5M-determinant CASCI the pipeline reaches 22.8 mHa, thus, a scalability demonstration limited by budget, not chemical accuracy.

The exact statevector simulation also runs across many nodes, each GPU holding an equal 34.4 GB slice; since adding one qubit doubles the state, each extra qubit doubles the GPU count. Measured this way (Table 2), the pipeline samples the full 40-qubit state, 8 TiB over 256 GPUs on 64 nodes, in 107.2 s per 5000-shot sample. Within one node extra circuit depth is cheap, but across nodes depth becomes much more expensive and a fixed per-sample overhead appears, so this route suits one-off exact validation of already-trained circuits, not training. Double precision at 40 qubits would need 512 GPUs, costly and beyond typical HPC partition limits, so the validated fp32 precision is required.

The pipeline is chemically accurate against exact references at every rung through 30 qubits, on SnO through 32, and on both charge states of the industrial n-butyl ligand. The open-shell methyl cation reaches 1.08 mHa at $L = 80$, from 1.84 to 1.90 at shallow depth.

The pipeline then ran on the real 54-qubit IQM Emerald processor (Table 3). At 14 qubits (SnO) it achieved +0.330 mHa, chemical accuracy on real hardware, with the identical pipeline at every stage, from the GPU simulator with native-gate emulation to the real device (42 logical CNOTs, 119 total logical gates, 5000 shots). At that rung the GEVP refinement was exact (0.0 mHa) in simulation. At 22 qubits (n-butyltin trihydroxide, the industrial ligand) the eight-circuit union of depth-truncated two-operator prefixes (8 to 12 logical CNOTs each, after a 13-CNOT probe chain collapsed to 2.4% modal fidelity; 40,000 shots total) reached +6.16 mHa with a single global readout remap, nearly 70% of the active-space correlation. Per-circuit readout calibration instead gives +3.92 mHa, 81% of that correlation, over the seven circuits that pass the validity gate. It reuses those same 40,000 shots and adds 40,000 calibration shots, and recovers the Hartree-Fock determinant in 7 of 8 circuits against 4 of 8 globally.

Table 4 collects the quantities a resist maker uses, with classical values beside them for the same bond. The 22q curve value is a difference of exact CASCI single points, which QSCI reproduces to chemical accuracy; the DFT and UCCSD(T) rows are full-molecule classical references for the separated fragments. Two limitations apply. The bond and ionisation energies combine quantum-verified points with classical references, and the neutral dissociation curve is a rigid stretch, both fragments frozen, so its bond energy is a lower bound. Figure 4 shows the switch itself. Six of the seven points along the Sn-

| Run | Qubits | vs CASCI | Detail |
|---|---|---|---|
| SnO GPU sim | 14 | +1.01 mHa | GEVP-refined: 0 mHa |
| SnO emulate | 14 | +0.227 mHa | native iqm target |
| SnO real QPU | 14 | +0.330 mHa | 54-qubit IQM, 5000 shots, 42 CNOT |
| n-Bu ligand, global remap | 22 | +6.16 mHa | 8 × 5000 shots, 3172-det union |
| n-Bu ligand, per-circuit cal. | 22 | +3.92 mHa | 81% of AS correlation |

Table 3: QSCI on the real 54-qubit IQM Emerald processor via CUDA-Q's native target. Hardware systems are SnO at 14q and the industrial n-butyl ligand at 22q; no $CH_3Sn(OH)_3$ hardware result is reported here. SnO (2e,7o) CASCI (−288.10912104 Ha); both SnO hardware rows within chemical accuracy. The 14q circuit is an eight-operator CCSD-amplitude-ordered pool prefix, not a policy sample; the 22q rows use depth-truncated two-operator prefixes of trained circuits, where per-circuit readout self-calibration is the accuracy step.

C stretch at 22 qubits are chemically accurate on the better of two seeds (Table 4). They stay accurate as the breaking bond turns diradical, its $\sigma^*$ natural occupation rising from 0.018 at equilibrium to 0.168 at 3.4 Å; the single miss is the most stretched point, 3.6 Å at 1.72 mHa, where the active-space descriptor is about to dissolve. That is the static-correlation regime in which density functional theory is known to fail [7], and where our DFT bond energy is 8 kcal/mol off. Removing one electron then weakens the bond 3.4-fold at UCCSD(T) (Figure 4 a), and the bond-energy and ionisation-energy cycle closes consistently. The same collapse appears for the real ligand, n-butyltin trihydroxide, with one twist. The n-butyl radical is easier to ionise than the tin fragment (7.35 against 7.48 eV at UCCSD(T)), so the charge leaves with the alkyl group, where for methyl it stayed on the tin side. The methyl radical's UCCSD(T) ionisation energy falls 0.30 eV below the evaluated experimental 9.843 eV [32], so this 0.13 eV channel-ordering margin is indicative rather than established.

## 4. Discussion

Within one system, accuracy is set by how much of a rung's determinant space the sampled subspace covers (Subspace, Table 1). Qubit count is a poor proxy for the size of that space. The 30-qubit resist space holds 207k determinants against 2.9M for the 26-qubit one, and both are chemically accurate. At 40 qubits coverage becomes the binding limit, set by determinant support, which determinants the circuits reach, rather than by sampler fidelity. Two checks test the obvious alternatives. Replaying the eight trained circuits at $\chi = 64/128/256$ returns identical unions and energies to better than $10^{-9}$ Ha, and natural-orbital rotation barely changes the Hartree-Fock weight ($|c_0|^2 \approx 0.957$). So neither MPS truncation nor the orbital basis is to blame, and more GPUs do not buy accuracy. The depth lever that closes the gap below 40q is blocked there, because each MPS rollout already costs over an hour. So 40q stops at a budget-limited demonstration while the SnO ladder is chemically accurate at 32 qubits, and classical subspace recovery, which needs no deeper circuit, is the tested route past it (below).

| Quantity | Method | $CH_3Sn(OH)_3$ | n-$BuSn(OH)_3$ |
|---|---|---|---|
| Neutral Sn-C bond energy | classical, DFT (B3LYP) | 64.7 kcal/mol | 59.8 kcal/mol |
| | classical, UCCSD(T) | 72.6 kcal/mol | 70.9 kcal/mol |
| | classical, exact CASCI (22q, lower bound) | 73.3 kcal/mol | - |
| Cation Sn-C bond energy | classical, UCCSD(T) | 21.2 kcal/mol | 30.3 kcal/mol |
| Weakening on ionisation | classical, UCCSD(T) ratio | 3.4× | 2.3× |
| Ionisation energy (adiabatic) | classical, DFT (B3LYP) | 9.70 eV | 9.00 eV |
| | classical, UCCSD(T), vert. / adiab. | 10.40 / 9.71 eV | 9.90 / 9.11 eV |
| Sn-C stretch on ionisation | DFT geometries | 2.141 → 2.416 Å | 2.155 → 2.400 Å |
| Cation splits into | | $CH_3$ radical + charged $Sn(OH)_3$ | charged n-Bu + $Sn(OH)_3$ radical |
| QSCI vs exact CASCI, 22q | quantum (this work) | neutral curve 6/7 points CA, cation curve 7/11 | both charge states CA |

Table 4: The EUV switch quantified at 22 qubits, classical and quantum values side by side. DFT misses the neutral bond energy by 8 to 11 kcal/mol across the ligand series yet errs only 0.01 to 0.11 eV on the adiabatic ionisation energy, so the bond that matters is where correlation is needed. The 22q curve, which QSCI tracks to chemical accuracy at six of seven points, lands at the coupled-cluster value; its rise is a rigid-stretch lower bound and the close agreement is partly fortunate cancellation. UCCSD(T) rows are classical references for the separated fragments; the methyl ones all-electron, the n-butyl ones frozen-core, so the methyl and n-butyl values are not directly comparable. The n-butyl channel assignment rests on a 0.13 eV margin, indicative rather than established.

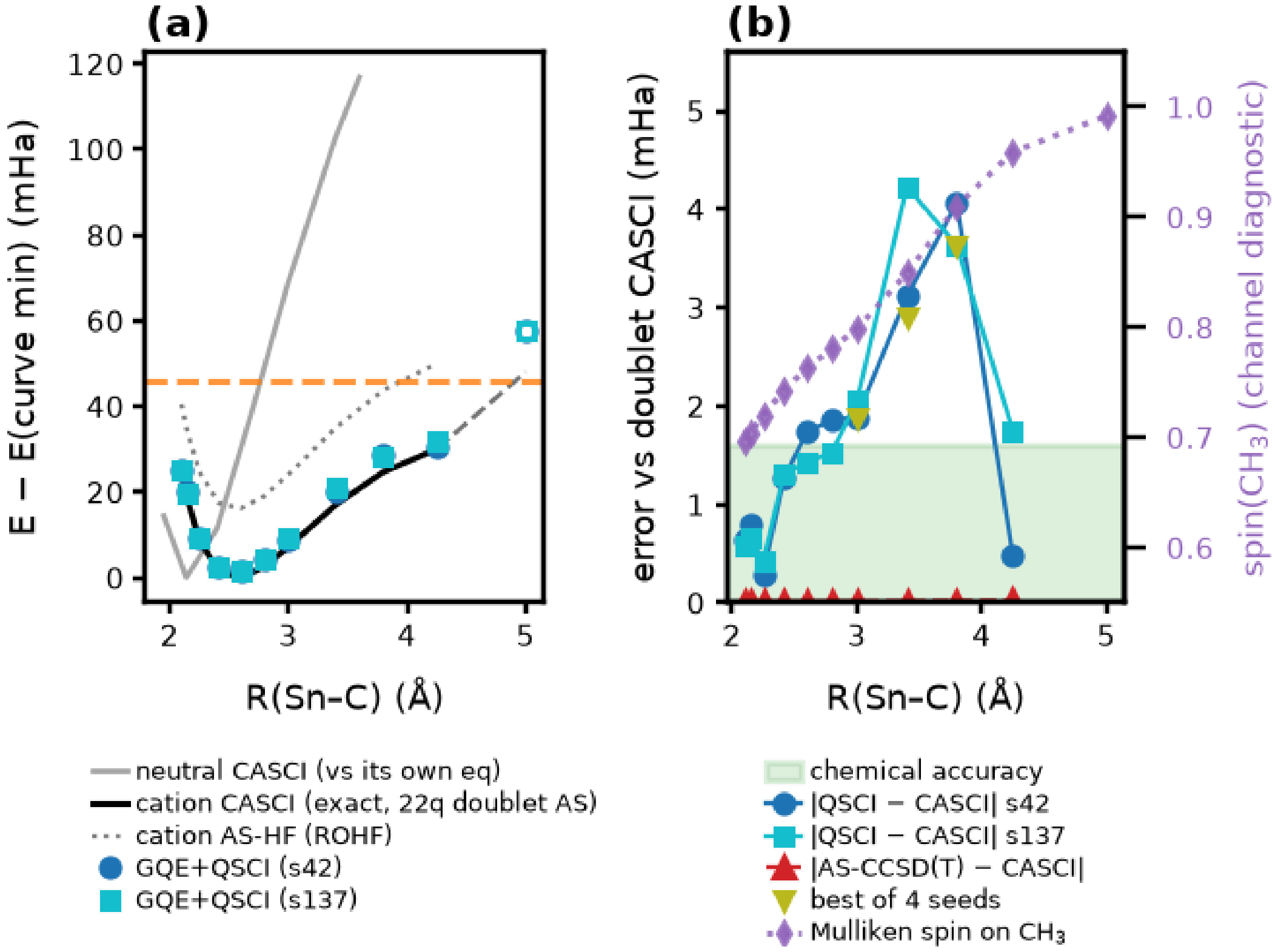

Figure 4: The EUV switch in $CH_3Sn(OH)_3$ at 22 qubits. (a) The radical cation's Sn-C dissociation surface: GEVP-refined QSCI energies ride the exact doublet CASCI curve out to 4.25 Å, two thirds of the way to the asymptote; the neutral curve (grey) is far steeper, and the dashed orange line is the $CH_3$ radical plus charged $Sn(OH)_3$ frozen-fragment gap at UCCSD(T). Ionisation collapses the classical UCCSD(T) bond-dissociation energy from 72.6 to 21.2 kcal/mol. (b) Per-point error against the chemical-accuracy band, with the active-space CCSD(T) deviation, the best of four seeds where run, and the Mulliken spin on $CH_3$ (right axis) tracking the channel. The open markers and dashed segment at 5.0 Å are the active-space boundary point, excluded from the curve claims; its error is 9.5 mHa, or 4.3 mHa after the orbital-ordering fix.

Exact verification also ends at 40q, because the AVAS spectrum jumps straight to 44q, whose FCI needs over 334 GB. The pipeline still trains there, with no exact reference to check it against, its 44q union recovering only 21.95 mHa of correlation energy, well short of the 75.1 mHa of a classical selected-CI bound and the 79.1 mHa of an active-space CCSD(T) estimate. Past 40 qubits we therefore have variational bounds and classical estimates rather than exact references, and the verified ladder stops there.

Device noise is a conditional feature for QSCI. It scattered the 22q circuits' one or two ideal determinants each into 60 to 203 particle-number-conserving strings per circuit, 862 in total (3172 determinants after spin completion); at 14q the emulate-to-hardware penalty was only +0.10 mHa with no error mitigation [16,17]. The same broadening appeared on the same device in a separate QSCI study of additive binding in asphalt, where it removed the need for zero-noise extrapolation [33]. To quantify that, noiseless sampling of the same truncated 22q circuits yields just two determinants (+19.45 mHa), so the hardware counts at +3.92 mHa improve on ideal sampling by 15.5 mHa. The same noise destroyed the full-depth run (an invalid +1433 mHa from uncorrected readout order and decoherence); the design response is the emulate-first gate, depth truncation, per-circuit readout self-calibration and persisted raw counts.

Every claim separates the solver gap (|QSCI − CASCI|, same active space) from the active-space gap (|CASCI − full correlated reference|), with classical baselines on the same instances. For the 14q demo RHF, CASCI, CCSD and CCSD(T) each run in under 3 s, and the 333 mHa CASCI-to-CCSD(T) difference is correlation outside the two-electron window.

The full-valence tin spaces are multireference. Active-space CCSD undershoots the exact CASCI correlation by 11.7 mHa at 16q, deepening to 15.5 mHa at 28 and 32q, so the cheap classical workhorse degrades exactly as the space grows (Table 5). The learned circuit reaches chemical accuracy on $N_2$ where the three-layer LUCJ circuit [34] fails at a comparable two-qubit gate count and a single Trotter step needs about $1.5 \times 10^4$ gates, while the refined subspace competes in compactness with heat-bath CI [14,18]. SKQD [35], a training-free quantum sampler reimplemented on the same rungs, is chemically accurate at every rung we ran it on through 26 qubits, so the learned sampler buys nothing there. From 28 qubits upward SKQD leaves the band at every rung we ran, while the learned sampler stays inside it on SnO through 32q and on the resist through 30q (Table 1). This is because SKQD's Trotter depth doubles with every Krylov power and its wall arrives two rungs earlier. Our own UCCSD-VQE and ADAPT-VQE runs [36–38] reach the same 14q instance's CASCI by driving a single deep ansatz through hundreds to thousands of energy evaluations. By contrast, the pipeline's best sampled circuit in simulation holds 44 CNOTs, 87 and 41 times fewer two-qubit gates, and its sampled union spans all 49 determinants, so the refinement returns the CASCI exactly. On published benchmarks GQE comes within 1.13 mHa of UCCSD-VQE with 58.8% fewer gates and 64.2% fewer CNOTs [39], and the generative approach moves all optimisation into the classical model, avoiding barren plateaus [40].

We implemented S-CORE, the self-consistent configuration recovery of Robledo-Moreno et al. [16], as a classical baseline (occupancy-weighted determinant search on the same Hamiltonian, engine cross-validated between PyCI and PySCF to $4 \times 10^{-10}$ mHa). Generative models have also been placed inside the recovery step itself, a restricted Boltzmann machine replacing SQD's configuration recovery [41], where ours instead emits circuits. Seeded from the pipeline's own budget-limited subspaces it drives the remaining exactly-referenced resist rungs, 32, 34 and 40q, into chemical accuracy. From a from-scratch Hartree-Fock seed with no GQE at all it still reaches 1.48 mHa at 32q and 1.25 mHa at 40q, the rung where the quantum sampler is budget-limited at 22.8 mHa (Table 1). We therefore claim no quantum-accuracy advantage at these classically tractable scales. Classical selected-CI solves these spaces, and no budget-matched seeded-versus-scratch comparison exists, so we claim no advantage from seeding either. This work instead establishes a hardware-ready pipeline verified against exact references at every rung through 40 qubits, a measured map of the coverage wall, and the demonstration that classical configuration recovery [16,17] is the route past it.

| Method, instance | Computation | Outcome |
|---|---|---|
| UCCSD-VQE, SnO 14q | quantum ansatz, simulated | reaches the CASCI; 3824 CNOTs × 490 evaluations |
| ADAPT-VQE, SnO 14q | quantum ansatz, simulated | 1808 CNOTs; 3112 evaluations with pool screening |
| LUCJ, $N_2$ 16q | quantum ansatz, simulated | 9.4 mHa at 808 gates; GQE precise at 786 |
| SKQD, rungs 14 to 32q | quantum sampling, simulated, no training, one seed | exact on the two-electron 14 and 18q rungs; chemically accurate through 26q; first miss at 28q |
| CCSD, SnO 16 to 32q | classical, single reference | undershoots exact CASCI by 11.7 to 15.5 mHa |
| Config. recovery, 14q qBraid run and 22q hardware | classical, on the QPU counts | +0.575 → +0.07; +3.92 → +0.18 mHa |
| S-CORE, resist 32 and 40q | classical, from scratch | 1.48 and 1.25 mHa with no quantum sampler |

Table 5: Baselines and levers on the instances used here; this work's own rows are Table 1 and Table 3. Rows compare only within their own instance. Quantum-simulated rows run quantum circuits on classical simulators; classical rows involve no quantum computation.

Four measured levers set the cost of the loop, and fitted scaling laws with a measured per-rung profile complete it (Table 6). The last is what the generative recipe buys, a single protocol reused unchanged and never retuned. That cost profile is dominated by compute rather than by the link to the QPU, which is why remote hardware access costs us little. An independent runtime model of hybrid workflows places both GQE and QSCI in the compute-bound regime, with communication-to-computation ratios of $10^{-4}$ to $10^{-1}$, and concludes that co-locating a QPU with the classical resource yields negligible benefit for workloads of this shape [42]. Our containerised sampling bridge targets the classical-side overhead that model leaves exposed. The same algorithm class has been run closed-loop at the opposite extreme of scale, a Heron processor with Fugaku at full scale [43]; the profile here is what that loop costs on one to 64 nodes.

| Lever or law | Measured effect |
|---|---|
| fp32 statevector | halves memory; exact 32q training on one A100, 75 to 89 min per 200-iteration seed |
| persistent containerised `mgpu` worker | steady-state rollout cost 61-fold lower at 24 qubits, 2.3-fold at 34; 34q trains at 25 to 50 min per $L = 600$ iteration on 4 pooled A100s |
| bounded-$\chi$ MPS | makes 40q sampling possible; moderate depth nearly free |
| protocol transfer | one unchanged training protocol across the 18 homolysis and cation-curve geometries; one unchanged 22q protocol across methyl, ethyl, n-propyl and n-butyl (1.42/1.27/0.66/0.72 mHa, seed 42) |
| sampling vs qubits | doubles per added qubit (exponent 0.995, $R^2 = 1.0$) |
| sampling vs depth | linear within one node ($R^2 = 0.998$) |
| diagonalisation | subspace dimension to the power 1.5 ($R^2 = 0.9999$) |
| 600-iteration run at $L = 600$ | about 8 GPU-hours per rung at 26 qubits, about 90 at 32 |

Table 6: Four measured levers on the cost of the training loop (upper block) and the cost model over the measured range (lower; the three scaling laws are fits, the GPU-hour figures measured). A companion HPC characterisation instruments the loop (per-phase timers, a GPU-hour cost model across qubits, depth, GPUs and backend) and supplies Table 2. 26 qubits is the largest resist rung chemically accurate on both seeds; the model estimates a new target's cost from its qubit count and intended depth.

Two entry points re-run the pipeline end to end by two routes to the same 14-qubit instance, each asserting its own exact CASCI reference before reporting. A pip-only mini GQE + QSCI loop on SnO (2e,7o)/14q is exact at +0.000 mHa in 16.1 s on a CPU-only host, all 49 determinants sampled. The hardware protocol then runs through the qBraid runtime, reaching +1.120 mHa on the free QIR simulator and +0.575 mHa from 31 of 49 determinants on the real 54-qubit Emerald (36,000 shots, all four readout maps bijective). That is an independent second route to the same instance's +0.330 mHa in Table 3. The real-QPU path is deliberately gated behind an explicit flag and a credit cap and priced before it runs, and the shipped raw counts replay that union offline with no key and no QPU spend. Across those two backends the sampled support widens under device noise, from 9 union determinants on the simulator to 31 on hardware, with both unions already chemically accurate. In both runs the readout convention was corrected blind, from the circuits' own counts alone, identifying the simulator's lexicographic bit order and the opposite convention on Emerald. Classical configuration recovery on the same counts, variationally safe as a strict superset [16], tightens the 14q union to +0.07 to 0.14 mHa and the 22q counts to +0.18 to 0.21 over three seeds each, a gain specific to these classically tractable subspaces. The shipped README maps each of these numbers to the command that reproduces it; a qBraid re-run reproduces the 14q pipeline and the scaling machinery in miniature, not the ladder itself.

Four limitations bound these claims. (i) At 40q the trained subspace is support-limited, so 22.8 mHa is a demonstration, not chemical accuracy. (ii) MPS truncation changes only which bitstrings are sampled; the bond-dimension ladder above, on that eight-circuit subset, shows no distortion of the 40q result, so the binding limit is (i). (iii) Classical recovery reaches the 32 to 40-qubit spaces without the quantum sampler, so that boundary is mapped, not beaten. (iv) On the QPU, routed two-qubit depth caps execution at shallow prefixes today, and the full trained ansatz awaits better fidelities. Classical confusion-matrix mitigation [44] of the persisted 14q counts (per-qubit rates from the shipped calibration variants, held-out validated, cross-checked against Mitiq's REM [45] to $10^{-15}$) leaves the union unchanged at matched support, at +0.575 mHa with the same 31 determinants. At every threshold it stays within 0.08 mHa. The same fit on the 22q counts gives no robust change, its rates dominated by circuit error at the depth wall. QSCI consumes counts only through determinant selection, so the residual error is missing sampling support, not miscounted weights, making classical recovery the effective lever.

## 5. Conclusion

A single CUDA-Q GQE + QSCI stack trains from 14 to 40 qubits against exact references at every rung, holding chemical accuracy through 30 qubits on the resist and 32 on SnO (best seed at each top rung). It runs the industrial resist ligand on a real 54-qubit QPU with self-calibrated readout and, with classical UCCSD(T) fragment anchors, quantifies the EUV solubility switch (72.6 → 21.2 kcal/mol on ionisation). The measured levers and the walls they run into (determinant support, cross-node depth cost, QPU routed depth) form a reproducible route for quantum materials informatics. Quantum-accurate descriptors can be delta-learned [46] across a ligand family. On the monoalkyltin series, a correction trained on three ligands cuts the DFT error on a fourth ligand's Sn-C bond energy 6.9-fold and its ionisation energy 4.0-fold (N = 4 proof of concept), pointing towards high-throughput resist design [47,48]. The route also complements the fault-tolerant road map. Quantum phase estimation needs high-overlap initial states, which mean-field references lose in exactly this multireference regime, and our compact QSCI subspaces are such states. The pipeline, both demonstrations and the persisted hardware counts are available at `github.com/KarimElgammal/gqe-qsci-euv-photoresists`.

## References

[1] D. Kazazis, J. Garcia Santaclara, J. van Schoot, I. Mochi, and Y. Ekinci, Extreme ultraviolet lithography, Nat. Rev. Methods Primers, DOI:10.1038/s43586-024-00361-Z **4**, 84 (2024).

[2] L. Li, X. Liu, S. Pal, S. Wang, C. K. Ober, and E. P. Giannelis, Extreme ultraviolet resist materials for sub-7 nm patterning, Chem. Soc. Rev., DOI:10.1039/c7cs00080d **46**, 4855 (2017).

[3] T. Manouras and P. Argitis, High Sensitivity Resists for EUV Lithography: A Review of Material Design Strategies and Performance Results, Nanomaterials, DOI:10.3390/nano10081593 **10**, 1593 (2020).

[4] J. H. Ma and others, Mechanistic Advantages of Organotin Molecular EUV Photoresists, ACS Appl. Mater. Interfaces, DOI:10.1021/acsami.1c12411 **14**, 5514 (2022).

[5] T. Du and others, Tin-oxo nanoclusters for extreme ultraviolet photoresists: Effects of ligands, counterions, and doping, J. Chem. Phys., DOI:10.1063/5.0200630 **160**, 154307 (2024).

[6] J. Li, Z. Wang, and H. Wang, Computational Study of Organotin Oxide Systems for Extreme Ultraviolet Photoresist, J. Phys. Chem. A, DOI:10.1021/acs.jpca.4c07585 **129**, 1420 (2025).

[7] A. J. Cohen, P. Mori-Sánchez, and W. Yang, Challenges for Density Functional Theory, Chem. Rev., DOI:10.1021/cr200107z **112**, 289 (2012).

[8] S. McArdle and others, Quantum computational chemistry, Rev. Mod. Phys. 92, 015003, DOI:10.1103/revmodphys.92.015003 (2020).

[9] T. D. Kharazi and others, Quantum Simulations for Extreme Ultraviolet Photolithography, arXiv:2602.20234 (2026).

[10] R. M. Cigala, C. De Stefano, A. Giacalone, A. Gianguzza, and S. Sammartano, Hydrolysis of Monomethyl-, Dimethyl-, and Trimethyltin(IV) Cations in Fairly Concentrated Aqueous Solutions at I = 1 mol L-1 (NaNO3) and T = 298.15 K. Evidence for the Predominance of Polynuclear Species, J. Chem. Eng. Data, DOI:10.1021/je101069y **56**, 1108 (2011).

[11] R. T. Frederick, S. Saha, J. T. Diulus, F. Luo, J. M. Amador, M. Li, D.-H. Park, E. L. Garfunkel, D. A. Keszler, and G. S. Herman, Thermal and radiation chemistry of butyltin oxo hydroxo: A model inorganic photoresist, Microelectron. Eng., DOI:10.1016/j.mee.2018.11.011 **205**, 26 (2019).

[12] B. Cardineau, R. Del Re, M. Marnell, H. Al-Mashat, M. Vockenhuber, Y. Ekinci, C. Sarma, D. A. Freedman, and R. L. Brainard, Photolithographic properties of tin-oxo clusters using extreme ultraviolet light (13.5 nm), Microelectron. Eng., DOI:10.1016/j.mee.2014.04.024 **127**, 44 (2014).

[13] NIST, *Tin Monoxide*, https://webbook.nist.gov/cgi/cbook.cgi?ID=C21651194&Mask=1000.

[14] R. Kemmoku, Q. Gao, S. Kanno, K. Keithley, I. Hamamura, N. Yamamoto, and K. Nakaji, Generative Circuit Design for Quantum-Selected Configuration Interaction, arXiv:2604.09756 (2026).

[15] M. Motta, W. Kirby, I. Liepuoniute, K. J. Sung, J. Cohn, A. Mezzacapo, K. Klymko, N. Nguyen, N. Yoshioka, and J. E. Rice, Subspace methods for electronic structure simulations on quantum computers, Electron. Struct. 6, 013001, DOI:10.1088/2516-1075/ad3592 (2024).

[16] J. Robledo-Moreno, M. Motta, and others, Chemistry Beyond the Scale of Exact Diagonalization on a Quantum-Centric Supercomputer, arXiv:2405.05068 (2024).

[17] N. Vaquero-Sabater, A. Carreras, L. Broers, T. Shirakawa, S. Yunoki, and D. Casanova, Noise and Configuration Recovery Impact on Quantum Selected Configuration Interaction, arXiv:2605.23697 (2026).

[18] A. A. Holmes, N. M. Tubman, and C. J. Umrigar, Heat-Bath Configuration Interaction: An Efficient Selected Configuration Interaction Algorithm Inspired by Heat-Bath Sampling, J. Chem. Theory Comput., DOI:10.1021/acs.jctc.6b00407 **12**, 3674 (2016).

[19] K. Nakaji and others, The Generative Quantum Eigensolver (GQE) and its Application for Ground State Search, arXiv:2401.09253 (2024).

[20] Y. Alexeev and others, Artificial intelligence for quantum computing, Nat. Commun., DOI:10.1038/s41467-025-65836-3 **16**, 10829 (2025).

[21] K. Kanno and others, Quantum-Selected Configuration Interaction: Classical Diagonalization of Hamiltonians in Subspaces Selected by Quantum Computers, arXiv:2302.11320 (2023).

[22] NVIDIA CUDA-Q development team, *CUDA-Q: The Platform for Hybrid Quantum-Classical Computing, Version 0.14*, https://nvidia.github.io/cuda-quantum.

[23] A. Radford and others, *Language Models Are Unsupervised Multitask Learners*, https://cdn.openai.com/better-language-models/language_models_are_unsupervised_multitask_learners.pdf.

[24] G. D. Purvis and R. J. Bartlett, A full coupled-cluster singles and doubles model: The inclusion of disconnected triples, J. Chem. Phys., DOI:10.1063/1.443164 **76**, 1910 (1982).

[25] M. Richer and others, PyCI: A Python-scriptable library for arbitrary determinant CI, J. Chem. Phys., DOI:10.1063/5.0219010 **161**, 132502 (2024).

[26] Z. Shao and others, DeepSeekMath: Pushing the Limits of Mathematical Reasoning in Open Language Models, arXiv:2402.03300 (2024).

[27] C. C. J. Roothaan, Self-Consistent Field Theory for Open Shells of Electronic Systems, Rev. Mod. Phys., DOI:10.1103/revmodphys.32.179 **32**, 179 (1960).
[28] J. R. McClean and others, OpenFermion: the electronic structure package for quantum computers, Quantum Sci. Technol. 5, 034014, DOI:10.1088/2058-9565/ab8ebc (2020).
[29] NIST, *Hydroxyl Radical*, https://webbook.nist.gov/cgi/cbook.cgi?ID=C3352576&Mask=1000.
[30] E. R. Sayfutyarova and others, Automated Construction of Molecular Active Spaces from Atomic Valence Orbitals, J. Chem. Theory Comput., DOI:10.1021/acs.jctc.7b00128 **13**, 4063 (2017).
[31] Q. Sun and others, Recent developments in the PySCF program package, J. Chem. Phys. 153, 024109, DOI:10.1063/5.0006074 (2020).
[32] NIST, *Methyl Radical, Gas Phase Ion Energetics*, https://webbook.nist.gov/cgi/cbook.cgi?ID=C2229074&Mask=20.
[33] K. Elgammal and M. Maußner, Additive binding energies in asphalt on a quantum processor via quantum-selected configuration interaction (QSCI), arXiv:2605.27640 (2026).
[34] M. Motta and others, Bridging physical intuition and hardware efficiency for correlated electronic states: the local unitary cluster Jastrow ansatz for electronic structure, Chem. Sci., DOI:10.1039/d3sc02516k **14**, 11213 (2023).
[35] J. Yu et al., Quantum-Centric Algorithm for Sample-Based Krylov Diagonalization, arXiv:2501.09702 (2025).
[36] A. Peruzzo and others, A variational eigenvalue solver on a photonic quantum processor, Nat. Commun., DOI:10.1038/ncomms5213 **5**, 4213 (2014).
[37] J. Tilly and others, The Variational Quantum Eigensolver: A review of methods and best practices, Phys. Rep., DOI:10.1016/j.physrep.2022.08.003 **986**, 1 (2022).
[38] H. R. Grimsley and others, An adaptive variational algorithm for exact molecular simulations on a quantum computer, Nat. Commun., DOI:10.1038/s41467-019-10988-2 **10**, 3007 (2019).
[39] K. Keithley and others, Auger Spectroscopy via Generative Quantum Eigensolver: A Quantum Approach to Molecular Excitations, arXiv:2603.12859 (2026).
[40] J. R. McClean and others, Barren plateaus in quantum neural network training landscapes, Nat. Commun., DOI:10.1038/s41467-018-07090-4 **9**, 4812 (2018).
[41] K. S. V. Anurag, A. K. Patra, M. Mukherjee, and others, Bridging the NISQ and Fault-Tolerant Regimes: Generative-ML-Assisted Quantum Selected CI for Molecular Simulations, arXiv:2606.30551 (2026).
[42] P. Rao et al., A Performance Model for Hybrid Quantum-Classical Workflows, arXiv:2607.15426 (2026).
[43] T. Shirakawa, J. Robledo-Moreno, T. Itoko, V. Tripathi, K. Ueda, Y. Kawashima, and others, Closed-loop calculations of electronic structure on a quantum processor and a classical supercomputer at full scale, arXiv:2511.00224 (2025).
[44] Z. Cai and others, Quantum error mitigation, Rev. Mod. Phys. 95, 045005, DOI:10.1103/revmodphys.95.045005 (2023).
[45] R. LaRose and others, Mitiq: A software package for error mitigation on noisy quantum computers, Quantum, DOI:10.22331/q-2022-08-11-774 **6**, 774 (2022).
[46] R. Ramakrishnan and others, Big Data Meets Quantum Chemistry Approximations: The Δ-Machine Learning Approach, J. Chem. Theory Comput., DOI:10.1021/acs.jctc.5b00099 **11**, 2087 (2015).
[47] D. A. Kreplin and others, sQUlearn: A Python Library for Quantum Machine Learning, IEEE Software, DOI:10.1109/ms.2025.3527736 **42**, 65 (2025).
[48] S. Minami and others, Generative Quantum Combinatorial Optimization by means of a novel Conditional Generative Quantum Eigensolver, Digital Discovery, DOI:10.1039/d5dd00138b **4**, 2229 (2025).